\documentstyle[prl,aps,epsf]{revtex}         

\begin{document}
%
%
%
\twocolumn[\hsize\textwidth\columnwidth\hsize\csname @twocolumnfalse\endcsname

\title{Comment on the paper:  Quantum backaction of optical observations on 
Bose-Einstein condensates by U. Leonhardt, T. Kiss, and P. Piwnicki, Eur. Phys. J. D7, 413 (1999)}
\author{Wolfgang Ketterle}
\address{Department of Physics and Research Laboratory of
Electronics, \\ Massachusetts Institute of Technology, Cambridge, MA 02139}
\date{\today}
\maketitle
\begin{abstract}

\end{abstract}
\pacs{PACS numbers: to be added} \vskip1pc ]
\narrowtext

Dispersive imaging with off-resonant light is an important technique for observing Bose-Einstein
condensates \cite{andr96,kett99var}. Compared to absorption imaging it causes much less heating,
and hence, allows the recording of non-destructive real-time ``movies'' of the dynamics of
Bose-Einstein condensates \cite{andr97prop}. We observed that a limitation of dispersive imaging
comes from residual absorption or Rayleigh scattering.  The momentum transfer to the condensate
atoms depletes the condensate and heats the cloud due to the transferred recoil energy
\cite{andr96,kett99var}.

In contrast, a recent paper \cite{leon99} emphasized that the limit of dispersive imaging is not
residual absorption, but a different form of quantum backaction of the probe light which was
determined with a new approach to quantum-optical propagation. This note points out that these
conclusions are incorrect, and that Rayleigh scattering is the dominant quantum backaction of
dispersive imaging.

First, the absorption rate cannot be completely suppressed by imaging with far-detuned light.  For
a desired signal-to-noise ratio, a further detuning has to be compensated by higher laser intensity
in such a way that the rate of far-wing absorption is constant \cite{kett99var}. The absorption
rate per atom is simply the Rayleigh scattering rate $ \gamma_s=\Gamma f_{\rm exc}$, where $\Gamma$
is the natural linewidth and the excited state fraction $f_{\rm exc} = (\omega_R/2\Delta)^2$ is
given by the Rabi frequency $\omega_R $ of the probe light and its detuning $\Delta$.  The recoil
due to the scattering of photons knocks atoms out of the condensate and depletes it with a rate
$\gamma_s$.

Leonhardt et al.\ \cite{leon99} derived an expression for the depletion of the condensate
$\gamma_L$ (their Eq. (62)).  The rate $\gamma_L$ turns out to be proportional to the absorption
rate $\gamma_s$ but is smaller by a factor of (3/16) \cite{leon-pc}. This indicates that the
calculated backaction is related to Rayleigh scattering.  It seems that it is just Rayleigh
scattering with the smaller prefactor caused by approximations of the theory.  Therefore, the
statement by the authors that their result is \emph{qualitatively} different from Rayleigh
scattering is inconsistent with their results.

Another major result of Ref.\ \cite{leon99} is that the phase diffusion rate is always smaller than
the depletion rate.  Our experiments \cite{andr96,andr97prop} were not sensitive to perturbations
of the phase, and we didn't estimate this effect.

In conclusion, residual absorption or Rayleigh scattering is the dominant perturbation of
dispersive imaging, and this process is the dominant quantum backaction of the probe light on the
Bose-Einstein condensate.

\bibliographystyle{prsty}

\begin{thebibliography}{1}

\bibitem{andr96}
M. Andrews, M.-O. Mewes, N. van Druten, D. Durfee, D. Kurn, and W. Ketterle,
  Science {\bf 273},  84  (1996).

\bibitem{kett99var}
W. Ketterle, D. Durfee, and D. Stamper-Kurn,  in {\em Bose-Einstein
  condensation in atomic gases}, {\em Proceedings of the International School
  of Physics Enrico Fermi, Course CXL}, edited by M. Inguscio, S. Stringari,
  and C. Wieman (IOS Press, Amsterdam, 1999), pp.\ 67--176.

\bibitem{andr97prop}
M. Andrews, D. Kurn, H.-J. Miesner, D. Durfee, C. Townsend, S. Inouye, and W.
  Ketterle, Phys. Rev. Lett. {\bf 79},  553  (1997).

\bibitem{leon99}
U. Leonhardt, T. Kiss, and P. Piwnicki, Eur. Phys. J. D {\bf 7},  413  (1999).

\bibitem{leon-pc}
U. Leonhardt, T. Kiss, and P. Piwnicki, reply to this comment.

\end{thebibliography}

\end{document}